# Domain-Specific Fine-Tuning and Prompt-Based Learning: A Comparative Study for developing Natural Language-Based BIM Information Retrieval Systems

Han Gao[a], Timo Hartmann[b], Botao Zhong[a*], Kai Li[a], Hanbin Luo[a]

a. National Center of Technology Innovation for Digital Construction, Huazhong University of Science & Technology, Wuhan, Hubei, China

b. Technische Universität Berlin, Gustav-Meyer-Allee 25, 13355, Germany

## Abstract

Building Information Modeling (BIM) is essential for managing building data across the entire lifecycle, supporting tasks from design to maintenance. Natural Language Interface (NLI) systems are increasingly explored as user-friendly tools for information retrieval in Building Information Modeling (BIM) environments. Despite their potential, accurately extracting BIM-related data through natural language queries remains a persistent challenge due to the complexity use queries and specificity of domain knowledge. This study presents a comparative analysis of two prominent approaches for developing NLI-based BIM information retrieval systems: domain-specific fine-tuning and prompt-based learning using large language models (LLMs). A two-stage framework consisting of intent recognition and table-based question answering is implemented to evaluate the effectiveness of both approaches. To support this evaluation, a BIM-specific dataset of 1,740 annotated queries of varying types across 69 models is constructed. Experimental results show that domain-specific fine-tuning delivers superior performance in intent recognition tasks, while prompt-based learning, particularly with GPT-4o, shows strength in table-based question answering. Based on these findings, this study identify a hybrid configuration that combines fine-tuning for intent recognition with prompt-based learning for question answering, achieving more balanced and robust performance across tasks. This integrated approach is further tested through case studies involving BIM models of varying complexity. This study provides a systematic analysis of the strengths and limitations of each approach and discusses the applicability of the NLI to real-world BIM scenarios. The findings offer insights for researchers and practitioners in designing intelligent, language-driven BIM systems. Future research should focus on enhancing scalability and improving model robustness in large-scale BIM projects.

# 1 Introduction

BIM represents a comprehensive digital representation of both the physical and functional attributes of a building [1]. It is defined as a digital engineering approach that not only facilitates information management but also supports collaboration throughout a building's entire lifecycle [2]. With the advancement of mobile internet and cloud computing, a wide array of BIM-enabled applications has emerged, enabling on-site access to building data for use cases such as construction verification [3], indoor navigation [4], building performance monitoring [5], and facility maintenance management [6].

The proliferation of mobile and web-based BIM platforms has significantly expanded the range of end-users, extending beyond architects and engineers to include maintenance personnel, energy managers, and property stakeholders. Due to limited formal training in BIM model manipulation and interpretation, many of these users interact with project data through simplified interfaces provided by mobile applications. Consequently, conventional interaction paradigms, which are typically optimized for domain experts using desktop-based authoring tools, are inadequate in supporting the information needs of this wider user base.

Natural Language Interfaces (NLIs) have emerged as a promising solution to lower the barrier between non-expert users and structured building information. Inspired by the success of AI-driven virtual assistants like Microsoft Cortana, Apple Siri, and Amazon Alexa [7], as well as ChatGPT [8]. Several studies [9–12] have explored the application of NLIs in BIM environments, demonstrating the potential of such approaches to enhance accessibility and usability. However, achieving accurate and context-aware information retrieval through natural language remains challenging due to the semantic complexity and domain-specific variability inherent in BIM data. Previous research has investigated both rule-based and machine learning–based techniques for interpreting natural language queries in BIM. Early approaches employed conventional Natural Language Processing (NLP) methods, including dependency parsing, named entity recognition, and template-based matching [13,14] [15,16]. While effective under constrained linguistic conditions, these methods are limited in handling the diversity of user expressions and ambiguity in BIM element names. More recent advances leverage pretrained transformer-based models, such as BERT [17] and GPT [8], which offer greater contextual understanding and generalization capabilities.

Among these, two dominant approaches have emerged: domain-specific fine-tuning, where general-purpose language models are adapted using task-specific training data; and prompt-based learning, which employs manually designed prompts to guide the model's response without retraining. Although both approaches have been preliminarily explored in developing BIM NLI systems [18–22], no systematic comparison has been undertaken to evaluate their relative effectiveness for BIM information retrieval tasks. In particular, there remains a lack of understanding regarding how such strategies perform across query types of different complexities, and how the effectiveness of the approaches can vary when applied to BIM models in different scales.

To address these gaps, this study adapts a classic two-stage natural language processing (NLP) framework for database question answering (DBQA) to the context of BIM information retrieval, and conducts a comparative analysis of two implementation strategies. A comparative analysis is conducted to evaluate different implementation approaches. The framework consists of two components: intent recognition and table-based question answering. Within this structure, we conduct a comparative evaluation of two model development approaches—domain-specific fine-tuning and prompt-based learning. Both strategies are evaluated on a manually annotated BIM query dataset and further assessed across BIM models of varying scale and complexity to examine their generalizability.

The contributions of this study are as follows: (1) This study conducts a systematic comparison of domain-specific fine-tuning and prompt-based learning approaches for BIM natural language interaction, with a focus on exploring their performance across diverse query types; (2) This study evaluates the generalizability of the proposed two-stage framework across BIM models of varying scales and complexities, and examines how the two strategies can be effectively combined within this framework. The analysis includes a discussion of hybrid configurations and provides insights into their practical applicability for real-world BIM scenarios.

The remainder of this paper is structured as follows. Section 2 reviews related work on natural language interfaces in BIM and recent developments in large language models. Section 3 outlines the proposed methodology and details the two model development strategies. Section 4 describes the dataset, presents experimental results, and compares the performance of different approaches. Section 5 evaluates the optimal model configuration across BIM models of varying complexity. Section 6 provides a discussion of the key findings, and Section 7 concludes the paper.

## 2 Related works

**2.1 BIM information retrieval**

BIM information retrieval refers to accessing and extracting relevant data and information from BIM models [23]. BIM serves as a comprehensive database encompassing building data, including text descriptions, geometric representations, and spatial and topological relationships [24]. The Industry Foundation Classes (IFC) data model has emerged as a widely adopted standard for exchanging BIM data [25]. Initially developed by the International Alliance for Interoperability (IAI) and currently maintained by buildingSMART [26], the IFC model embodies an open, neutral, and non-proprietary format that facilitates information exchange between various BIM software applications [27]. Comprehensive access to diverse types of information within the IFC becomes crucial to develop a system capable of supporting BIM information retrieval. This includes basic attribute information, spatial relationships, and user-defined attributes. In the IFC framework, building information is primarily expressed through instances, employing objects such as IfcProject, IfcSite, IfcBuilding, IfcBuildingStorey, IfcSpace, IfcBuildingElement, and other IFC classes [27].

Several studies have developed BIM information retrieval approaches grounded in information derived from the IFC model. BIMQL represents an early attempt of BIM information retrieval. This approach employs several keywords such as "where," "select," "set," and "relationship" to specify conditional query statements that retrieve a single BIM element or element set that meets the query statement [28]. QL4BIM is another endeavor focused on defining a BIM query language, offering support for spatial queries by incorporating topological operators to define filter expressions with spatial semantics [29]. BIMRLSS was introduced as another simplified query language to enhance data accessibility in BIM models, transforming BIM data into a simplified schema that minimizes complex queries and allows highly optimized queries to be performed [30]. In order to further enable queries for semantic information in BIM models, some researchers have proposed ontology-based query approaches, where ontology links heterogeneous data from BIM and other sources, and queries are performed using the SPARQL query language. This approach facilitates the incorporation of domain concepts into BIM information retrieval applications [31].

Recognizing the difficulties in mastering computer query languages, some studies turn to visual programming languages (VPL) to facilitate BIM information retrieval. VPL offers a more intuitive and user-friendly interface for BIM data retrieval [32–34]. For example, Ghannad et al. parameterized the

VPL query rules into categories of target building elements, attributes, geometry, relationships, and logical expressions. The designer can transform design rules into a graphic interface to perform information retrieval [34].

From these studies, a notable trend emerges towards developing user-friendly interfaces for querying information in BIM models. However, these tools also require the user to master programming knowledge or specialized query rules, which creates a barrier for a broader range of potential users.

**2.2 Traditional NLP techniques in BIM information retrieval**

In the past few years, there has been increasing interest in incorporating NLP tools into BIM platforms. Several studies have explored the use of NLP to enable information retrieval in BIM systems, aiming to realize the interaction between natural language queries and the BIM database.

The early attempts employed traditional NLP techniques such as keyword mapping, syntax parsing, and sentence template matching in designing BIM information retrieval approaches. For example, Lin et al. proposed a method based on keyword mapping for IFC classes. This method allowed users to retrieve specific BIM elements by simply inputting relevant keywords, which were then matched to the predefined IFC categories [35]. Similarly, Wang et al. developed a natural language-based BIM information retrieval system that used part-of-speech tagging and syntactic analysis to match queries with the IFC database at the word level [14]. However, these methods had notable limitations, as word-level mapping struggled to capture semantic relationships between words, often resulting in inaccurate or incomplete retrieval results.

To enhance semantic matching, ontology-augmented NLP has been introduced. For example, Wu et al. employed a semantic query engine that mapped extracted phrases to IFC-related ontology concepts [16], and Yin et al. proposed a semantic parser that converted queries into executable code using ontological reasoning [13]. While ontology support improved conceptual alignment, these systems still struggled with variations in user phrasing and lacked scalability for reasoning tasks.

**2.3 Prompt-based learning and domain-specific fine-tuning**

Recent advancements in deep learning have led to the adoption of transformer-based pretrained language models, such as BERT [17] and GPT [36], which offer contextualized language understanding and generalization across diverse query structures. These models have shown promise in BIM-related applications through two main adaptation strategies: prompt-based learning and domain-specific fine-tuning.

In prompt-based learning, pretrained large language models (LLMs) are guided using carefully constructed textual prompts, eliminating the need for retraining. For instance, there have been several proof-of-concept applications that employ prompt-based learning on LLMs to enable interaction between BIM and the user [19–22]. More broadly, prompt-based interfaces have shown promise in interpreting building code for compliance checking [37], information searching from technical documents [38], and operation and maintenance management [39]. Prompt-based approaches offer flexibility and scalability, allowing non-expert users to interact with complex data sources through natural language without the overhead of dataset labeling or model retraining.

On the other hand, domain-specific fine-tuning involves adapting a pretrained model such as such as BERT and its derivatives [40–42] to a particular task or dataset by further training it on domain-relevant examples. This strategy allows the model to internalize specialized terminologies, contextual patterns, and semantic nuances unique to the target field, thereby improving performance on downstream applications.

While prior studies have proposed proof-of-concept applications of prompt-based learning for BIM information retrieval, its efficiency relative to domain-specific fine-tuning remains unclear. Several gaps persist in evaluating the effectiveness of these approaches. First, there is a lack of comparative studies examining their performance across queries of varying complexity. Additionally, the scalability and robustness of current BIM information retrieval approaches under different BIM model scales have not been assessed.

## 3 Methodology

### 3.1 Comparative study design

To investigate and compare domain-specific fine-tuning and prompt-based learning approaches for BIM natural language interaction, this study establishes a structured evaluation workflow, as illustrated in **Figure 1**. First, a generic two-stage natural language processing (NLP) framework is adapted for the BIM information retrieval context, within which both modeling approaches are implemented and evaluated. To support the evaluation, we construct a dedicated dataset that includes queries of different complexities. Using this dataset, we evaluate and compare the performance of both approaches across diverse query categories. Subsequently, we identify the optimal model configuration for the two-stage framework and assess its robustness and generalizability by applying it to BIM models of varying scales and complexities through a case study.

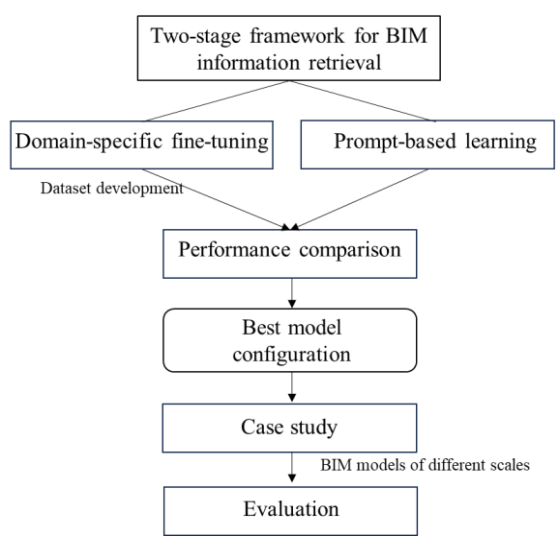

**Fig. 1.** Methodological workflow for the comparative study

### 3.1 Two-stage framework for BIM information retrieval

This study adopts a two-stage NLP framework for information retrieval in BIM. As illustrated in **Figure 2**, the framework consists of two core components: intent recognition, which classifies user queries into predefined categories, and table-based question answering, which retrieves relevant information from structured BIM data.

BIM models consist of different sub-databases that represent different types of building elements (e.g., spaces, doors, windows, columns). To support information retrieval, user queries are routed to the corresponding sub-database. In the preprocessing step, building element data is extracted and categorized into sub-databases, each labeled according to the corresponding element class. The intent recognition component is designed to identify the semantic type of each user query and assign it to the relevant element class. This enables the system to select the correct sub-database for downstream processing. Following this, a table-based question answering component is implemented to interpret the user query and return a relevant answer from the selected sub-database.

To implement this framework, we explore two different modeling approaches. First, a domain-specific fine-tuning approach is adopted, in which both components of the system are trained on the domain-specific dataset. Specifically, XLNet is used for intent recognition, while TAPAS is employed for answer generation over structured tabular data. Second, a prompt-based learning approach is adopted, which leverages large language models (LLMs) such as GPT series without additional training by designing task-specific prompts.

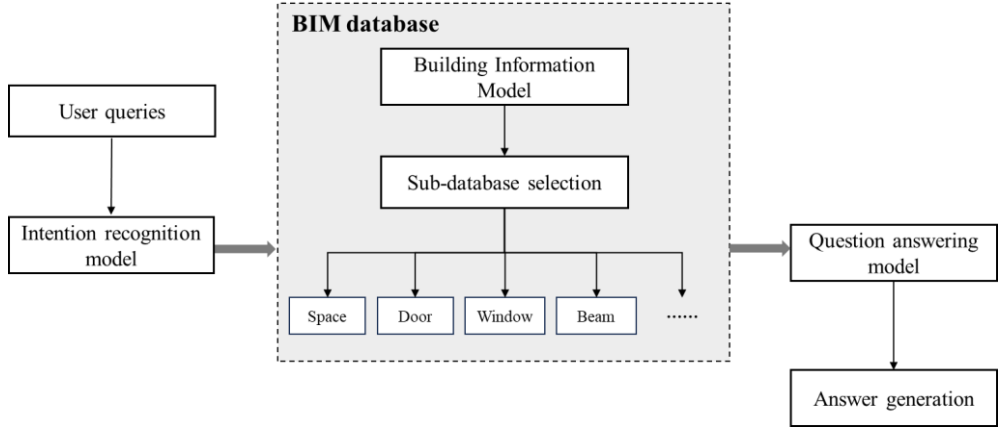

Fig. 2. Two-stage framework for natural language-based BIM information retrieval

### 3.2 Domain-specific fine-tuning approach
### 3.2.1 XLNet-based intent recognition model

The intent recognition model is developed to perform multi-label classification for user queries, enabling it to establish correlations between input user queries and relevant BIM sub-databases. For example, as shown in **Figure 3**, when a user inputs a question such as "How many doors are there on Level 2?", the intent recognition model can understand that the user is querying about door information and matches the user query with the sub-database extracted from the IfcDoor instances in the BIM model.

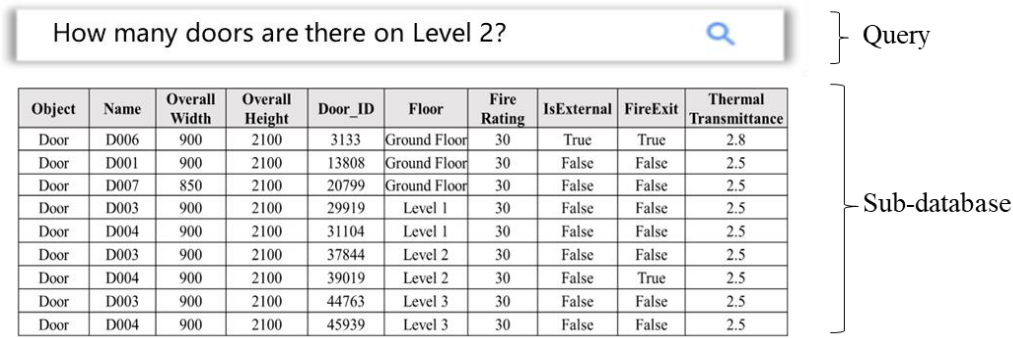

Fig. 3. An example of matching a user query (about door information) to the corresponding sub-database

The intent recognition model needs to handle different query statements and match them with sub-databases. It is built upon a transformer-based architecture called XLNet [43]. XLNet integrates a relative encoding scheme and a segment recurrence mechanism from Transformer-XL to capture long dependencies in sentences. It is pre-trained on a massive corpus of text data, which enables it to learn a wide range of language patterns and knowledge from the text. It excels in capturing contextual information and relationships within text.

We extend the XLNet architecture by adding a fully connected layer and a multi-label classifier to realize intent recognition for user queries. **Figure 4** illustrates the architecture of the intent recognition model.

The input user query is tokenized and converted into numerical representations, and input features are added as the embedding layer, including segment encodings and positional encodings. Two-stream self-attention is the core part of the XLNet architecture, enabling the model to capture relationships between tokens in both directions (i.e., left-to-right and right-to-left). This helps the model understand the context of each token within the entire sentence. The output from the self-attention layer is then passed through a fully connected layer to further transform and abstract the features learned from the input. The fully connected layers are followed by a multi-label classifier for predicting various labels(e.g., door, window, beam, among others) based on the contextual information learned from the input.

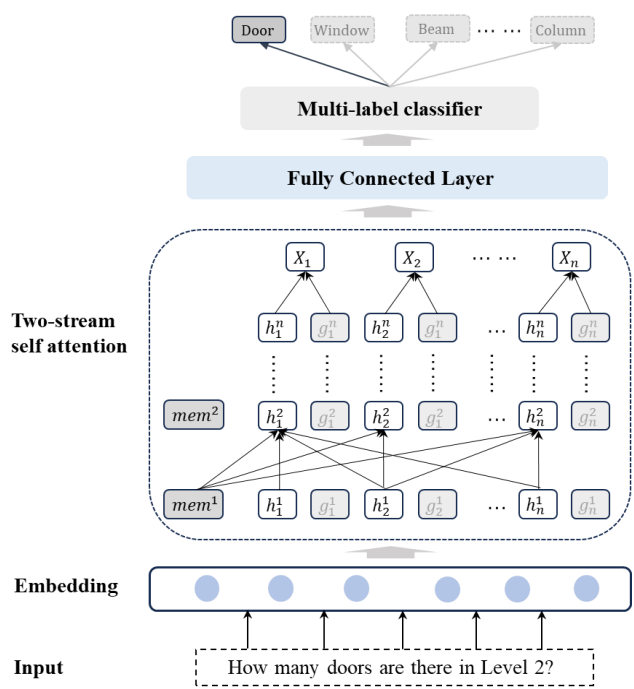

Fig. 4. The architecture of the intent recognition model based on XLNet

### 3.2.2 TAPAS-based table question answering model

This domain-specific fine-tuning employs the TAPAS architecture to train the table question-answering model to realize information retrieval on BIM sub-databases. The reason for choosing this architecture is as follows: First, TAPAS provides a pre-trained language model trained on the MLM (Masked Language Model) task with millions of tables from Wikipedia, enabling it to handle expression diversity in tabular data structures [44]. Second, TAPAS extends BERT's architecture [17] to encode tables as input, incorporating column embedding, row embedding, rank embedding, and introducing an aggregation prediction layer that allows it to handle queries involving comparative reasoning and aggregation operations.

As shown in **Figure 5**, the model transforms the input table and the user query into representations that capture both the table's structural information and the query's semantic meaning. For a query such as "Which are the top two floors with the largest number of doors?", the model outputs the cell selection and aggregation results, combining two results to generate an answer to the query. The model architecture consists of three major parts: the embedding layer, cell selection, and aggregation prediction.

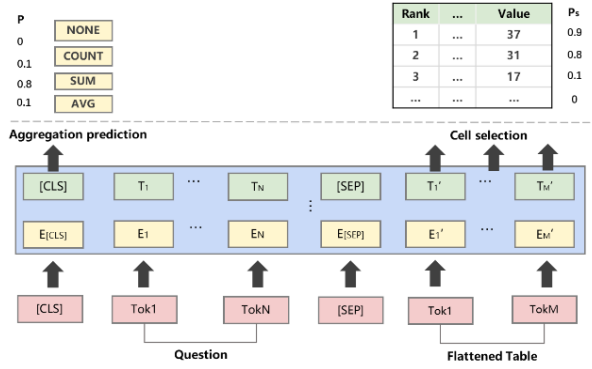

**Fig. 5.** Model architecture for table question-answering tasks

(1) Embedding Layer

The embedding layer includes token embedding, position embedding, segment embedding, column embedding, row embedding, and rank embedding. Based on the above encoding, the model can understand the table's structure and learn the relationship between questions and tables.

(2) Cell selection

The model chooses a subset of table cells and determines the appropriate aggregation operation to apply to them. When a single cell or a group of cells holds the answer, the model selects one or multiple cells accordingly. However, if the answer necessitates further aggregation operations like count, sum, or average, the model selects multiple cells and performs calculations on them to generate the answer. The loss function of the cell selection task consists of three components, as shown in Equation (4). $L_{cols}$ is the average binary cross-entropy loss over column selections, where the model selects the column with the most correlated cells in set C, as described in Equation (1). $L_{cells}$ is the binary cross-entropy loss for selecting cells within the selected column, as shown in Equation (2). $L_{aggr}$ is the logarithmic loss for aggregation. For cell selection tasks that do not involve aggregation operations, the aggregation value is defined as NONE, and the corresponding aggregation operation is assigned $op_0$, as shown in formula (3).

$$L_{cols} = \frac{1}{|Cols|} \sum_{co \in Cols} CE(p_{col}^{(co)}, \mathbb{I}_{co=col}) \quad (1)$$

$$L_{cells} = \frac{1}{|Cells\ (col)|} \sum_{c \in Cells(Col)} CE(p_s^{(c)}, \mathbb{I}_{c \in C}) \quad (2)$$

$$L_{aggr} = -\log\ p_a(op_0) \quad (3)$$

$$L_{CellSelection} = L_{cols} + L_{cells} + L_{aggr} \tag{4}$$

(3) Aggregation prediction

The answer not only involves directly selecting the cell values in the table but also involves selecting the aggregation operator, such as COUNT, SUM, AVERAGE, and NONE. Therefore, the model incorporates an additional linear classification layer to predict the aggregation operator. The softmax function is used to calculate the probability $p_a^{(op)}$ of selecting an output aggregation operator on the last hidden layer. The aggregation operator op is output when the output pa(op) is greater than 0.5.

### 3.2.3 Data annotation strategy for domain-specific fine-tuning

To enable supervised fine-tuning in a domain-specific context, the dataset was annotated to support both the intent recognition model and the table-based question answering model. **Figure 6** shows some annotated examples. Workers labeled the question, the address of the table, the coordinates of the answer in the table, the answer text, the aggregation label, the float answer, and the table label to enable the training process. The explanation for the annotation is as follows:

- question: The question is represented as a string. For example, "Can you let me know the window count inside the building?"
- table_file: This is the CSV file (BIM sub-database) address corresponding to the question asked.
- answer_coordinates: A list composed of tuples formed by the coordinates corresponding to the answer text in the table. Each tuple represents a row-column pair of the cells involved in the answer. For instance, ['(0, 6)', '(1, 6)', '(2, 6)', '(3, 6)'].
- answer_text: A list of strings or a single string.
- aggregation_label: The index of the aggregation operator. Used for supervised training of aggregation prediction. The aggregation operators such as NONE, SUM, AVG, and COUNT are represented by index values 0, 1, 2, and 3 respectively.
- float_answer: The float answer to the question. For example, when the answer text is ['92'], the float answer is 92.
- table_label: The label corresponding to the table, used for training the intent recognition model. Queries were labeled with corresponding information types, which include floor, space, window, door, beam, column, stair, and furniture.

| question | table_file | answer_coordinates | answer_text | aggregation_label | float_answer | table_label |
|---|---|---|---|---|---|---|
| Can you let me know the window count inside the building? | Savoyvilla_floor.csv | ['(0, 6)', '(1, 6)', '(2, 6)', '(3, 6)'] | ['92'] | 1 | 92 | floor |
| The number of windows in space the id of which is 321? | Simple3-storeytestbuilding_space.csv | ['(1,5)'] | ['7'] | 0 | | space |
| The long name of space which has more than 2 doors? | Simple3-storeytestbuilding_space.csv | ['(4,4)'] | ['Level 1 Room 2'] | 0 | | space |
| Which floor does not have any column? | 01_BIMcollab_Example_ARC_floor.csv | ['(0,1)'] | ['-01. Fundering'] | 0 | | floor |
| Which floor has most number of railing? | 01_BIMcollab_Example_ARC_floor.csv | ['(2,1)'] | ['01 verdieping'] | 0 | | floor |
| How many spaces are there in 02 verdieping? | 01_BIMcollab_Example_ARC_floor.csv | ['(4,9)'] | ['4'] | 0 | 4 | floor |
| The elevation of F2? | 12-1-1_floor.csv | ['(2,2)'] | ['3600'] | 0 | 3600 | floor |
| How many beams in F4? | 12-1-1_floor.csv | ['(4,10)'] | ['11'] | 0 | 11 | floor |
| The number of windows in which floor is 114? | 12-1-1_floor.csv | ['(2,1)'] | ['F2'] | 0 | | floor |
| Which floor have spaces? | 12-1-1_floor.csv | ['(1,1)','(2,1)'] | ['F1','F2'] | 0 | | floor |

**Fig. 6.** Examples of data annotation for domain-specific fine-tuning

### 3.3 Prompt-based learning approach

Prompt-based learning involves crafting specific input prompts that guide the model toward producing the desired output without requiring full model retraining. This approach is advantageous in scenarios where limited task-specific data is available or where fast adaptation to new query patterns is needed. We employ GPT series models (GPT-4 and GPT-4o) in the few-shot setting.

The prompts employed consist of three components: a role-play instruction, few-shot examples, and a task instruction. The role-play instruction is used to instruct the LLM to simulate an expert for BIM information retrieval. Few-shot examples give several examples to train the LLM to classify user queries and answer user queries based on the given BIM database information.

For the intent recognition task, the prompt structure is shown in **Figure 7**. Each example consists of a query and the label of the corresponding BIM subdatabase, which covers queries towards different subdatabases. The role-play instruction defines the model's role as a BIM assistant and outlines the expected label list. The few-shot examples provide query–response pairs, demonstrating how query intents map to BIM sub-database categories (e.g., "How many bathrooms are there in the building?" → space). Lastly, the task instruction presents a new query and asks the model to perform the query classification.

For the table question answering task, the prompt structure is shown in **Figure 8.** Each example consists of a query, relevant BIM database information, and a corresponding labeled answer. It covers query types of different complexities, including a simple query, a query involving comparative reasoning, and a query involving aggregation calculation. The role-play instruction defines the model's function as a BIM assistant specialized in extracting information from structured BIM data in tabular format. The few-shot examples demonstrate how the model is expected to interpret natural language questions and locate relevant values in the database. For example, a query like "What is the elevation of 01 eerste verdieping in relation to ground level?" is answered with ['0.0'], based on the provided spatial data table. The tabular database is presented in plain-text format and includes key-value pairs or row-based records representing BIM components and their attributes (e.g., levels, windows, elevations). The model is expected to learn both the semantic structure of the table and the question-answering logic from the provided examples. The task instruction then presents a new table and user query, prompting the model to retrieve the answer directly from the data. This format enables the model to perform structured reasoning and numeric lookup without requiring hardcoded rules or schema parsing.

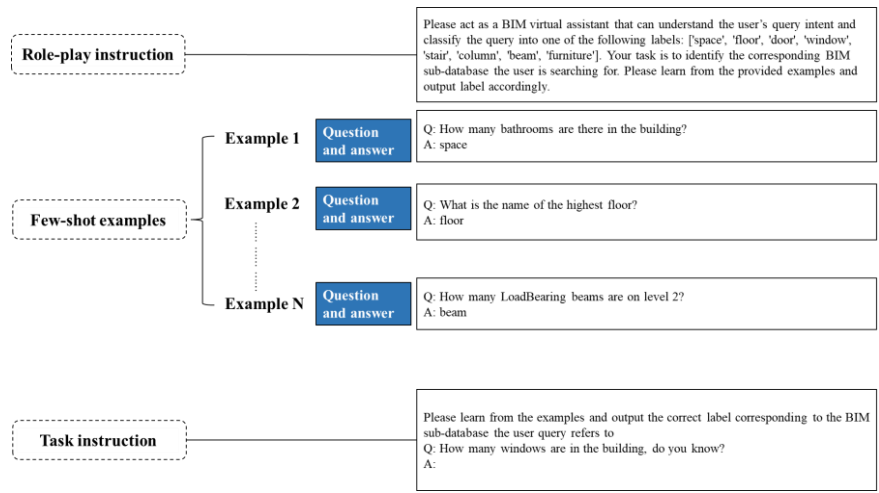

**Figure. 7.** Structure of prompts for intent recognition

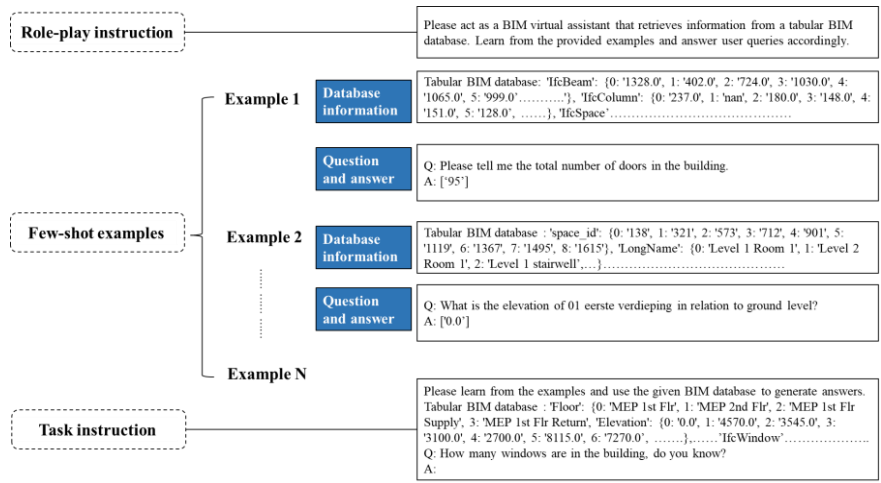

**Figure. 8.** Structure of prompts for table question answering

# 4 Comparative analysis of two different approaches

**4.1 Dataset Development**

This study developed 1740 BIM-related queries related to IfcStair, IfcSpace, IfcBeam, IfcColumn, IfcWindow, IfcDoor, and IfcFurniture over 69 BIM models. These queries were constructed based on 69 BIM models obtained from the Open IFC Model Repository [45]. The BIM tabular databases used to perform question answering are extracted from IFC files through predefined rules. In addition to covering a wide variety of BIM entities (e.g., doors, spaces, windows, beams), the dataset encompasses distinct query types that require varying levels of semantic understanding and reasoning. The typology includes queries that focus on element attributes, spatial relationships, comparative reasoning (e.g., maximum, minimum, greater than, less than), and aggregation operations (e.g., count, average, summation). The question-answering data was manually annotated by two trained annotators with background knowledge in BIM. **Table 1** summarizes the query types along with representative examples illustrating their typical usage.

Table 1 Query typology and representative examples in the developed dataset

| Query Type | Description | Representative Example |
|---|---|---|
| Attribute query | Retrieve specific property values of individual BIM elements | What is the width of all staircases on the second floor? <br> What is the fire rating of each internal door on Level 1? |
| Spatial relationship | Identify spatial or topological relationships between BIM components | Which windows belong to Room_205? <br> Which doors are connected to Space 102? |
| Comparative reasoning | Compare attributes across multiple elements to determine relative magnitudes | Which room on Level 2 has the largest floor area? <br> Which beam has the smallest cross-sectional area?" |
| Aggregation operation | Calculate statistical measures over a group of elements (e.g., count, sum, average) | How many doors are located on each floor? <br> What is the average length of windows on the fifth floor? |

**4.2 Experiment setting**

The dataset was randomly partitioned into two subsets: 80% for training and 20% for testing. The training set was exclusively used for model optimization, while the testing set remained unseen during training and was used solely for performance evaluation.

The domain-specific fine-tuned models were evaluated against a prompt-based learning baseline using GPT-4 and GPT-4o, where few-shot prompts were manually constructed and submitted via the official web interface of GPT. We fine-tuned two pretrained transformer-based architectures: XLNet for intent recognition and TAPAS for table-based question answering. All training experiments were conducted on a local workstation equipped with an Intel Core i5-13600KF processor (3.50GHz), 32GB RAM, and an NVIDIA RTX 4090 GPU with 24GB VRAM. The intent recognition model was optimized using the AdamW optimizer with a learning rate of 2e-5 and a batch size of 32, and was trained for 10 epochs. For the table question answering model, a similar setup was employed, with a learning rate of 5e-5, batch size of 32, and training for 15 epochs.

## 4.3 Performance comparison between domain-specific fine-tuning and prompt-based learning approaches

### 4.3.1 Overall Performance

This section presents an evaluation and comparison between domain-specific fine-tuning and prompt-based learning approaches across two tasks: intent recognition and table-based question answering. The assessment focuses on model performance across different query types, demonstrating their strengths, limitations, and complementarities. The results for all tested models are summarized in **Table 2**. The domain-specific fine-tuned model demonstrates significantly higher accuracy in intent recognition (99.14%) compared to both prompt-based solutions. In contrast, prompt-based learning with ChatGPT-4o achieves the highest performance in table-based question answering (91.95%), surpassing the fine-tuned TAPAS model (86.49%). However, GPT-4 performs notably worse in both tasks, with lower accuracy in intent recognition and table question answering compared to the domain-specific models. Notably, a hybrid configuration in which fine-tuned XLNet handles intent recognition and GPT-4o processes table question answering achieves the highest overall accuracy at 90.80%. This result suggests that combining specialized models for different tasks can lead to superior performance.

**Table 2** Performance comparison between prompt-based learning and domain-specific fine-tuning for intent recognition and table-based question answering

| Approach | Intent recognition | Table question answering | Overall accuracy |
| --- | --- | --- | --- |
| Prompt-based learning (GPT-4) | 43.97% | 76.15% | 34.77% |
| Prompt-based learning (GPT-4o) | 47.13% | **91.95%** | 41.09% |
| Domain-specific fine-tuning (XLNet-TAPAS) | **99.14%** | 86.49% | 85.75% |
| Hybrid Configuration (XLNet + GPT-4o) | **99.14%** | **91.95%** | **90.80%** |

### 4.3.2 Intent Recognition Performance

For the intent recognition task, which involves classifying user queries into predefined intent categories (space, floor, door, window, stair, column, beam, furniture), the fine-tuned XLNet model achieves a remarkably high accuracy of 99.14%, significantly outperforming the prompt-based learning approaches using GPT-4 and GPT-4o. **Figure 9** presents the confusion matrix, which illustrates the model's performance in classifying user intents by showing the alignment between true labels and predicted labels. It is important to note that the BIM sub-databases are not strictly exclusive in content; a table labeled with a specific element type can still contain references to other related entities in order to support spatial relationship queries. For example, a table labeled as "Door" may also include the IDs of spaces connected to each door. As a result, when a user poses a query such as "Which door is Space 40156 connected to?", prompt-based models often struggle to determine whether to retrieve information from the "Door" or the "Space" database. This ambiguity exists because prompt-based learning relies solely on few-shot examples and lacks domain-specific training. In contrast, the domain-specific XLNet model, having been trained on a substantial amount of annotated queries, is capable of learning these implicit associations. It can more accurately interpret the underlying intent and correctly map the query to the most relevant sub-database, even when information overlaps across different tables.

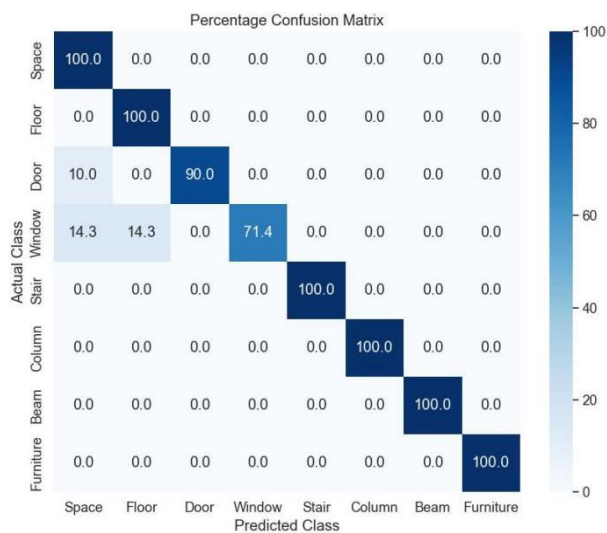

**Fig. 9.** Confusion matrix of the XLNet-based intent recognition model

**4.3.3 Table-Based Question Answering Performance**

We compared three settings for table-based question answering: prompt-based learning with GPT-4, prompt-based learning with GPT-4o, and a domain-adapted TAPAS model. Each setting was evaluated on four query types (attribute queries, spatial-relationship queries, comparative reasoning, and aggregation operations).

As is shown in **Figure 10**, prompted GPT-4o achieved the highest overall accuracy (91.95 %), followed by the domain-adapted TAPAS model (86.49 %) and prompted GPT-4 (76.15 %). By query type, prompted GPT-4o outperformed domain-adapted TAPAS on attribute queries, comparative reasoning, and aggregation operations, and matched domain-adapted TAPAS on spatial-relationship queries (96.35 % vs. 97.08 %). In contrast, domain-adapted TAPAS surpassed prompted GPT-4 on all query types except for comparative reasoning, demonstrating the value of domain adaptation for table understanding. These results indicate that there is no universally superior approach between domain-adapted and prompt-based models for table question answering; rather, the effectiveness of each strategy depends heavily on the specific version and capabilities of the underlying large language model.

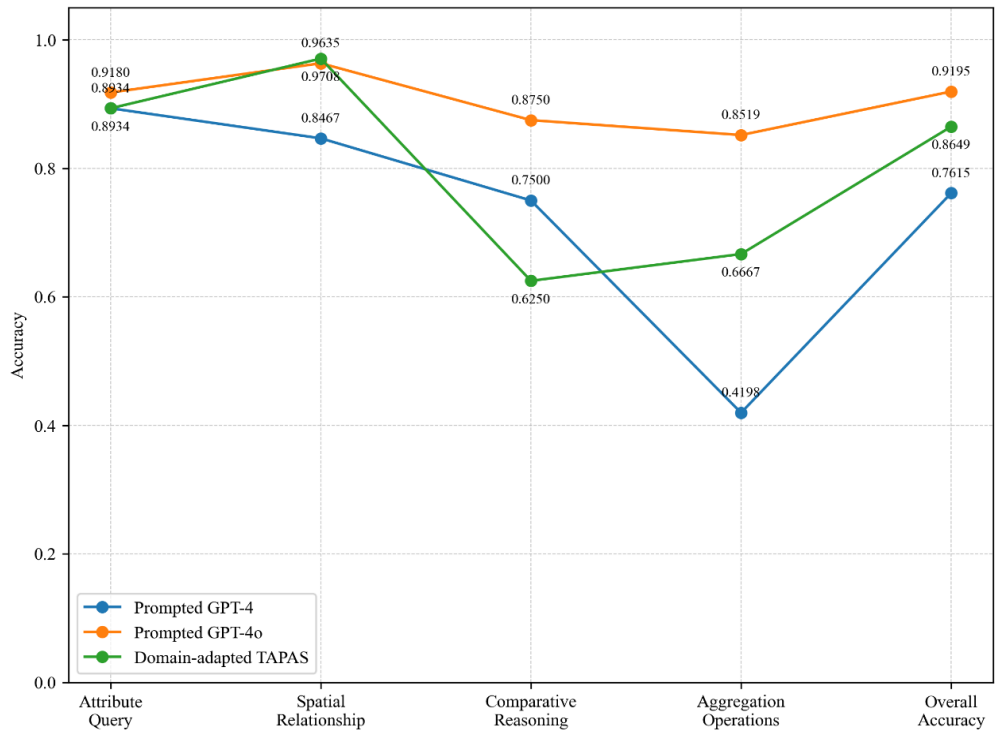

**Fig.10.** Accuracy comparison between prompt-based learning (GPT) and domain-specific fine-tuned model (TAPAS) for table-based question answering.

## 5 Case study

Following the comparative analysis, as shown in **Section 4.3.1**, the optimal model configuration for the proposed BIM information retrieval framework was identified as a combination of an XLNet-based intent recognition model and the prompted GPT-4o for table-based question answering. To evaluate the generalizability of this configuration, a case study was conducted involving BIM models of varying complexity.

Three BIM models representing different building scales were selected for the study: a small-scale detached two-story house with 7 spaces, a medium-scale office building with 4 floors and 14 spaces, and a large-scale teaching facility comprising 5 floors and 91 spaces. These models, illustrated in **Figure 11**, reflect increasing levels of architectural and data complexity.

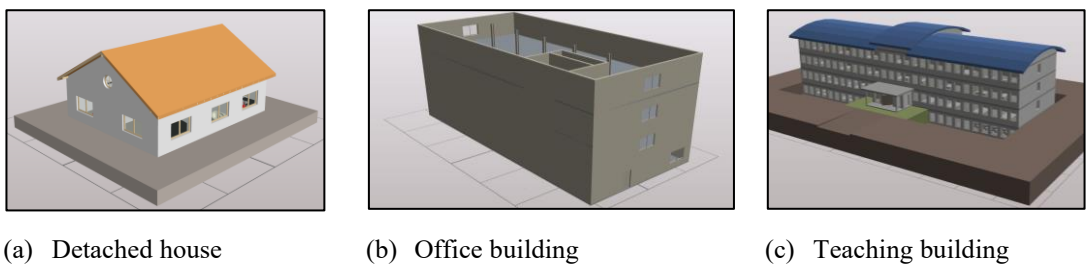

(a)  Detached house　　　　　(b)  Office building　　　　　(c)  Teaching building

**Fig. 11.** BIM models of three case buildings with varying levels of complexity.

Two participants were asked to generate ten queries for each building, the queries maintained consistent syntactic structures across models and encompassed three categories of tasks: descriptive (e.g., retrieving floor area), comparative (e.g., identifying elements based on attribute thresholds), and aggregative (e.g., computing the total quantity of specific elements). The test results for each case are shown in **Table 3, 4, and 5.** The record of GPT-4o's responses to the question-answering tasks is included in the **Appendix A.** In Case 1 (the small-scale model) and Case 2 (the medium-scale model), the framework achieved perfect performance by successfully classifying all intents and retrieving correct answers for all queries. However, in Case 3 (the large-scale model), while the intent recognition module maintained full accuracy, the performance of the GPT-4o-based question answering component deteriorated. Specifically, the model failed to correctly answer 6 out of 10 queries. One of these errors involved a complete failure in an aggregation task: for the query "What is the total number of doors in the building?", the model generated an entirely incorrect total. For four other failed queries, the answers were incomplete subsets of the correct results. This performance degradation is attributed primarily to the scale of the input tables. The largest subdatabase in Case 3 contained over 200 rows, which exceeded the effective context window and processing capacity of the GPT-4o model.

This limitation stems from the inherent characteristics of transformer-based architectures, where the self-attention mechanism exhibits quadratic time and space complexity with respect to input length. As a result, the model's ability to process long sequences diminishes significantly when applied to large-scale structured data. In such scenarios, the model tends to return partial results, failing to comprehensively traverse or interpret the entire dataset. This behavior is consistent with that observed in other transformer-based table models like TAPAS, which also demonstrate a tendency to return incomplete subsets of the correct answer when faced with extensive tabular inputs.

Overall, despite the high accuracy achieved on smaller BIM models, the optimal configuration experiences a notable decline in performance when applied to large-scale projects. This underscores the need for future work focused on enhancing the scalability and robustness of LLM–based table question answering systems.

Table 3 Test results of different learning strategies on Case 1

|  | Intent recognition | Table question answering | Overall Results |
|---|---|---|---|
| How many windows are there in space 211242? | True | True | True |
| Which spaces have a GrossFloorArea larger than 20? | True | True | True |
| What is the LongName of the space that has a height greater than 3 meters? | True | True | True |
| Id of spaces that have 2 windows? | True | True | True |
| what is the total number of doors in the building? | True | True | True |
| What is the ID of the door that has a height greater than 2.2 meters? | True | True | True |
| What are the names of doors that have a width less than 0.9 meter? | True | True | True |
| What are the IDs of the windows in the building with a height of less than 1.1 meters? | True | True | True |
| Which space is window 17435 located in? Please output the ID of the space. | True | True | True |
| Find the windows on the Dachgeschoss floor and output their IDs. | True | True | True |

Table 4 Test results of different approaches on Case 2

|  | Intent recognition | Table question answering | Overall Results |
|---|---|---|---|
| How many windows are there in space 211242? | True | True | True |
| Which spaces have a GrossFloorArea larger than 20? | True | True | True |
| What is the LongName of the space that has a height greater than 3 meters? | True | True | True |
| Id of spaces that have 2 windows? | True | True | True |
| what is the total number of doors in the building? | True | True | True |
| What is the ID of the door that has a height greater than 2.2 meters? | True | True | True |
| What are the names of doors that have a width less than 0.9 meter? | True | True | True |
| What are the IDs of the windows in the building with a height of less than 1.1 meters? | True | False | False |
| Which space is window 17435 located in? Please output the ID of the space. | True | True | True |

| | Intent recognition | Table question answering | Overall Results |
|---|---|---|---|
| Find the windows on the Dachgeschoss floor and output their IDs. | True | False | False |

Table 5 Test results of different approaches on Case 3

| | Intent recognition | Table question answering | Overall Results |
|---|---|---|---|
| How many windows are there in space 211242? | True | False | False |
| Which spaces have a GrossFloorArea larger than 20? | True | False | False |
| What is the LongName of the space that has a height greater than 3 meters? | True | True | True |
| Id of spaces that have 2 windows? | True | False | False |
| what is the total number of doors in the building? | True | False | False |
| What is the ID of the door that has a height greater than 2.2 meters? | True | False | False |
| What are the names of doors that have a width less than 0.9 meter? | True | True | True |
| What are the IDs of the windows in the building with a height of less than 1.1 meters? | True | True | True |
| Which space is window 17435 located in? Please output the ID of the space. | True | True | True |
| Find the windows on the Dachgeschoss floor and output their IDs. | True | False | False |

# 6 Discussion

This study presents a comparative evaluation of two approaches: domain-specific fine-tuning and prompt-based learning, for developing natural language interfaces to support information retrieval in BIMs. The experimental results and case study collectively offer several insights regarding their performance and limitations.

In the intent recognition task, the domain-specific fine-tuning approach demonstrates stronge, where it achieves very high accuracy. This can be attributed to its exposure to task-specific annotations and domain-relevant vocabulary during training, allowing it to effectively learn the syntactic and semantic nuances of queries, as well as the mappings between queries and corresponding sub-databases. In contrast to domain-specific fine-tuning, prompt-based learning operates in a few-shot setting without direct exposure to enough training examples. As a result, it struggles to disambiguate overlapping labels. When two or more table labels co-occur in one query, the few-shot learning fails to determine which sub-database the user is referring to, leading to low performance in intent recognition.

Conversely, in table-based question answering, prompted GPT-4o outperforms the fine-tuned TAPAS model, particularly in handling complex reasoning tasks such as comparative and aggregation queries. This suggests that recent general-purpose large language models possess sufficient latent reasoning capabilities to operate effectively even without domain-specific fine-tuning, provided the task complexity remains within their processing limits. However, the performance differences between GPT-4 and GPT-4o highlight the sensitivity of prompt-based strategies, whose effectiveness depends mainly on the specific model version. While the most advanced large language models deliver better performance, their high deployment cost makes domain-specific models a viable alternative for applications.

The case study further tests the optimal combination between the domain-specific fine-tuning approach and the prompt-based learning approach under different BIM model complexities. While the optimal configuration performs perfectly on small and medium BIM models, a substantial performance drop is observed in the large-scale case. Here, even GPT-4o fails to generate correct answers for 60 percent of the queries. The root cause lies in the transformer architecture's inability to handle long tabular contexts, particularly when tables exceed several hundred rows. Thus, even state-of-the-art LLMs are not yet robustly scalable for large tabular databases. This result is important for researchers and industry practitioners considering LLM-based BIM assistants, as it highlights that current solutions must be applied with caution, particularly in large-scale BIM projects.

The contribution of this research lies in articulating a quantitative comparison between domain-specific and prompt-based NLP approaches in the context of BIM information retrieval. While prior studies have explored the use of large language models (LLMs) for question answering, few have critically examined their strengths and limitations in comparison to domain-specific approaches, particularly across a diverse typology of BIM queries using a large-scale dataset. This study highlights the trade-offs and synergies between domain-specific fine-tuning and prompt-based learning for BIM information retrieval. While each approach has strengths in different sub-tasks, achieving robust, scalable performance in real-world BIM applications will likely require a fusion of both paradigms.

Nonetheless, both approaches exhibit limitations that must be addressed in future work. Domain-specific fine-tuning requires annotated data and substantial computational resources. Prompt-based learning, though more flexible and faster to deploy, often suffers from inconsistent performance across different model versions and incurs high operational costs when using advanced LLMs. Moreover, both approaches face scalability issues when processing large-scale tabular data in BIM models.

A potential solution to the challenges associated with processing large-scale tabular data involves the automatic partitioning of oversized tables, particularly those exceeding a predefined threshold of columns or rows, into smaller and more manageable segments. For example, tables can be divided at regular intervals, such as every 30 rows, with each segment processed independently by the language model. Intermediate answers generated for each segment can then be aggregated based on the query type, such as comparative or aggregation queries, to construct the final response. Although this segmented approach offers a potential means of addressing input length limitations, it requires multiple rounds of inference and subsequent integration of partial results, which may introduce cumulative errors. The effectiveness and reliability of this method should be further evaluated across a range of large-scale BIM models. Due to current data limitations, this evaluation could not be conducted in the present study and is therefore left for future work.

Beyond the scope of BIM information retrieval, future research could also explore how the two-stage framework could be extended to support more advanced BIM applications, such as automated building code compliance checking and language-guided BIM modeling. Realizing this potential will require not only stronger integration between LLMs and BIM schemas but also new methods for reasoning over regulatory logic, spatial constraints, and domain-specific knowledge in a verifiable and explainable manner.

# 7 Conclusion

This study presents a comparative analysis of domain-specific fine-tuning and prompt-based learning for natural language interaction with BIM. By establishing a domain-specific BIM query dataset and employing a two-stage framework that includes intent recognition and table-based question answering, we evaluated the strengths and limitations of both approaches.

Our results reveal that domain-specific fine-tuning delivers superior accuracy in intent recognition, benefiting from exposure to domain-specific annotations. In contrast, prompt-based learning, especially when using advanced models such as GPT-4o, demonstrates strong reasoning capabilities in table-based question answering, particularly for complex aggregation and comparative queries. The optimal configuration, combining fine-tuned XLNet for intent recognition with GPT-4o for table question answering, achieves high performance on the test dataset.

However, case studies highlight scalability challenges. The hybrid approach's performance degrades notably when applied to the large-scale BIM model, due to input length constraints inherent to transformer architectures. This finding highlights the limitations of LLM-based approaches in achieving accurate BIM information retrieval within large-scale project contexts.

Overall, this research contributes to the advancement of natural language processing strategies for BIM information retrieval. The findings provide insights for the development of NLI for BIM-based platforms, particularly in selecting appropriate NLP approaches based on project scale and task complexity. Future work should focus on enhancing scalability and contextual reasoning to support more adaptable systems.

# Acknowledgment

This research is partly supported by National Key R&D Program of China (2022YFC3801700), National Natural Science Foundation of China (Grant No. 72271106, No. U21A20151).

# Appendix A

**Records of GPT-4o Question Answering Performance on Case 1-3**

**Case 1**
Query 1: How many windows are there in space 211242?
GPT-4o: In space 211242, there are 2 windows. True

Query 2: Which spaces have GrossFloorArea larger than 20?
GPT-4o: The spaces with a GrossFloorArea larger than 20 are: Wohnen - 24.87555 m²
Schlafzimmer - 22.0725 m² Galerie - 107.16 m² True

Query 3: What is the LongName of the space that has a height greater than 3 meters?
GPT-4o: The space with a height greater than 3 meters is Galerie, and its LongName is 7.
False

Query 4: Id of spaces that have 2 windows?
GPT-4o: The spaces with 2 windows and their corresponding space IDs are: 210718 210837 210972 211242 299678 True

Query 5: What is the total number of Doors in the building?
GPT-4o: The total number of doors in the building is 8. True

Query 6: What is the ID of the door that has a height greater than 2.2 meters?
GPT-4o: From the data provided, the door with a height greater than 2.2 meters is the "Terrassentuer" with a height of 2.375 meters. Its door_id is 16149. True

Query 7: What are the names of doors that have a width less than 0.9 meter?
GPT-4o: The names of doors with a width less than 0.9 meters are: Innentuer-2 Innentuer-3 Innentuer-1 True

Query 8: What are the IDs of the windows in the building with a height of less than 1.1 meters?
GPT-4o: The windows with a height less than 1.1 meters are:
Window ID 223746 (row 9)
Window ID 236183 (row 10)
These correspond to OG-Fenster-1 and OG-Fenster-2, each with a height of 1.0 meters. True

Query 9: Which space is window 17435 located in? Please output the ID of the space?
GPT-4o: Window 17435 is located in the space with ID 302939. True

Query 10: Find the windows on the Dachgeschoss floor and output their IDs.
GPT-4o: The windows located on the Dachgeschoss floor have the following IDs:
223746
236183.
True

**Case 2**

Query 1: How many windows are there in space 16987?

GPT-4o: In space 16987, there are 2 windows, as indicated in the column "number of windows". True

Query 2: Which spaces have GrossFloorArea larger than 200?

GPT-4o: The spaces with a GrossFloorArea larger than 200 are:

Space ID 16987 (Office) - 277.02

Space ID 32793 (Office) - 295.52

Space ID 40706 (Office) - 295.52

Space ID 47269 (Office) - 295.52

These are all the spaces meeting the criteria. True

Query 3: What is the LongName of the space that has a height greater than 2500?

GPT-4o: Space ID 41494 LongName: Corridor True

Query 4: Id of spaces that have 2 windows?

GPT-4o: From the data provided, the spaces with 2 windows are:

space_id: 16987

space_id: 32793

space_id: 40706

space_id: 47269 True

Query 5: The total number of doors in the building?

GPT-4o: The total number of doors in the building is 16. True

Query 6: What is the ID of the door that has a width less than 880?

GPT-4o: The ID of the door that has a width less than 880 is 20799. True

Query 7: What are the names of doors that have a height more than 2500?

GPT-4o: The names of the doors that have a height greater than 2500 are D003 and D009. True

Query 8: What are the IDs of the windows in the building with a height of less than 1300 millimeters?

GPT-4o: In the provided dataset, all windows have a height of 1500 millimeters, which is greater than 1300 millimeters. Therefore, there are no windows with a height less than 1300 millimeters in this building. True

Query 9: Which space is window 36859 located in? Please output the ID of the space?
GPT-4o: Window with ID 36859 is located in the space with space0_id = 53639 True

Query 10: Find the windows on the ground floor and output their IDs.
GPT-4o: The windows located on the Ground Floor have the following IDs:
1960
4314
True

**Case 3**
Query 1: How many windows are there on space 35102?"
GPT-4o: Space 35102 has 2 windows. True

Query 2: Which spaces have GrossFloorArea larger than 20?"
GPT-4o: The IDs of the spaces with a GrossFloorArea larger than 20 are: 34757, 34932, 35102, 35252, 35422, 35598. False (incomplete)

Query 3: What is the LongName of the space that has a height greater than 3 meters?"
GPT-4o: There are no spaces in the data with a height greater than 3 meters. Let me know if you'd like to analyze other aspects of the dataset! True

Query 4: Id of spaces that have 4 windows?
GPT-4o: The IDs of the spaces that have exactly 4 windows are: 34757, 34932, 65361, 65516, 65842, 67921. Let me know if you want details about these spaces or further analysis. False (incomplete)

Query 5: What is the total number of Doors in the building?
GPT-4o: The total number of doors in the building is 95. Let me know if you want to analyze doors per floor, room type, or any other category. False (Incorrect, correct answer is 153)

Query 6: What are the IDs of the doors that are wider than 1.8?
GPT-4o: The ID of the door that is wider than 1.8 meters is 44321. Let me know if you'd like more details about this door or others based on different criteria. False (incomplete)

Query 7: What are the names of doors that have a height more than 2.2?
GPT-4o: The names of the doors that have a height greater than 2.2 meters are: Tür-016, Tür-002. True

Query 8: What are the IDs of the windows that have a height of less than 1.3 meters?

GPT-4o: There are no windows in the dataset with a height less than 1.3 meters. All windows are taller than that. Let me know if you want to filter by other height ranges or explore other attributes. True

Query 9: Which space is window 4071 located in? Please output the ID of the space?
GPT-4o: The window with ID 4071 is located in the space with ID 139358. Let me know if you want more information about this space or related elements. True

Query 10: Find the windows on the floor Keller and output their IDs.
GPT-4o: The IDs of the windows located on the Keller floor are: 3903, 4071, 4220, 4368, 4516, 4664, 4812, 4960, 5109, 5257, 5405, 5553, 5843, 5991, 6140, 6288, 6436, 6584, 6732, 6880, 7029, 7177, 7325, 7473, 8106. False (Incomplete)